\documentclass[showpacs,showkeys,aps]{revtex4}

\usepackage{amssymb}
\usepackage{amssymb}
\usepackage{amsmath}
\usepackage{graphicx}
\usepackage{dcolumn}

\newcommand{\ee}[1]{\mathrm{e}^{#1}}
\newcommand{\dd}{\mathrm{d}}
\newcommand{\ii}{\mathrm{i}}

\begin{document}

\title{Method to measure off-axis displacements based on the analysis of the
intensity distribution of a vortex beam}

\author{G. Anzolin}
\email{gabriele.anzolin@unipd.it}
\altaffiliation[Also at ]{INAF - Osservatorio Astronomico di Capodimonte,
salita Moiariello 16, I-80131 Napoli, Italy}
\author{F. Tamburini}
\author{A. Bianchini}
\author{C. Barbieri}
\affiliation{Dipartimento di Astronomia, Universit\`{a} di Padova,
vicolo dell'Osservatorio 3, I-35122 Padova, Italy}

\date{\today}

\begin{abstract}
We study the properties of the Fraunhofer diffraction patterns produced by Gaussian
beams crossing spiral phase plates. We show, both analytically and numerically, that
off-axis displacements of the input beam produce asymmetric diffraction patterns.
The intensity profile along the direction of maximum asymmetry shows two different
peaks. We find that the intensity ratio between these two peaks decreases exponentially
with the off-axis displacement of the incident beam, the decay being steeper for higher
strengths of the optical singularity of the spiral phase plate. We analyze how this
intensity ratio can be used to measure small misalignments of the input beam with a
very high precision.
\end{abstract}

\pacs{42.25.Bs, 42.25.Gy, 42.90.+m, 42.79.-e}

\keywords{Wave optics, Optical vortices, Phase singularities}

\maketitle

\section{Introduction}
\label{sec:intro}

Optical vortices (OVs) appear in light beams carrying screw wavefront dislocations
(vortex beams) \cite{nye74}. The surface of constant phase of a vortex beam has an
helical structure and presents phase singularities endowed with topological charge.
Beams harboring OVs carry also a quantity of orbital angular momentum (OAM)
\cite{all92} associated to the precession of the Poynting vector around the vortex
axis \cite{pad95}.

OVs have attracted an increasing interest in applied physics
\cite{gah96,mol01,cur03,gri03,mar05,mol07} and also for astronomical applications
\cite{swa01,har03,thi07,anz08,eli08,ber08}. In fact, they can be
easily produced in light beams with the help of specific optical devices that have a
central optical singularity. Among these optical elements, the most efficient ones are
fork holograms (FHs) \cite{baz90} and spiral phase plates (SPPs) \cite{bei94}.
Laguerre-Gaussian (L-G) modes have been often used to describe the beams produced
with such devices. However, a more precise description of the diffraction patterns
produced by an SPP \cite{ber04} or a FH \cite{sac98,bek08a} is provided by
hypergeometric (Kummer) functions. We shall use this approach in this Paper.

Consider an input beam with an amplitude distribution symmetric about the propagation
axis. When such a beam intersects an SPP or a FH perpendicularly and exactly on-axis,
it produces a circularly symmetric beam with a central dark region, where the field
amplitude is zero. Any misalignment with respect to the central discontinuity would
then produce an asymmetry of the observed intensity distribution \cite{vaz02} and the
topological charge of the correspondent off-axis OV may have a non-integer value
\cite{oem04a}. This changes also the OAM originally carried by the beam \cite{mai01},
thus producing an OAM spectrum \cite{vas03,vas05}.

The sensitivity of a vortex beam to displacements of the input beam has been
proposed as an indicator of nanometric shifts in a speckle pattern \cite{wan06}
or to be used as a non-interferometric method for the correction of small surface
deviations on spatial light modulators \cite{jes07}. Similar results could be
obtained from the analysis of the mean square value of the resulting OAM spectrum
\cite{liu08}. We also proposed a method to measure very small displacements based
on the degree of asymmetry of the intensity pattern of an off-axis vortex beam
\cite{tam06,anz08}. However, detailed analytical studies of the actual structure
of off-axis OV produced with SPPs of FHs were initiated only very recently
\cite{bek08b}. In this Paper, we extend the analysis of the intensity distribution
of an off-axis vortex beam generated under Fraunhofer diffraction conditions.
We then derive a more convenient formalism of our method for the detection of off-axis
displacements, in view of future applications with optical imaging devices.

\section{Fraunhofer diffraction of a Gaussian beam intersecting a spiral phase plate
on-axis} \label{sec:gaussdiff}

\begin{figure}
\begin{center}
\includegraphics[width=\columnwidth]{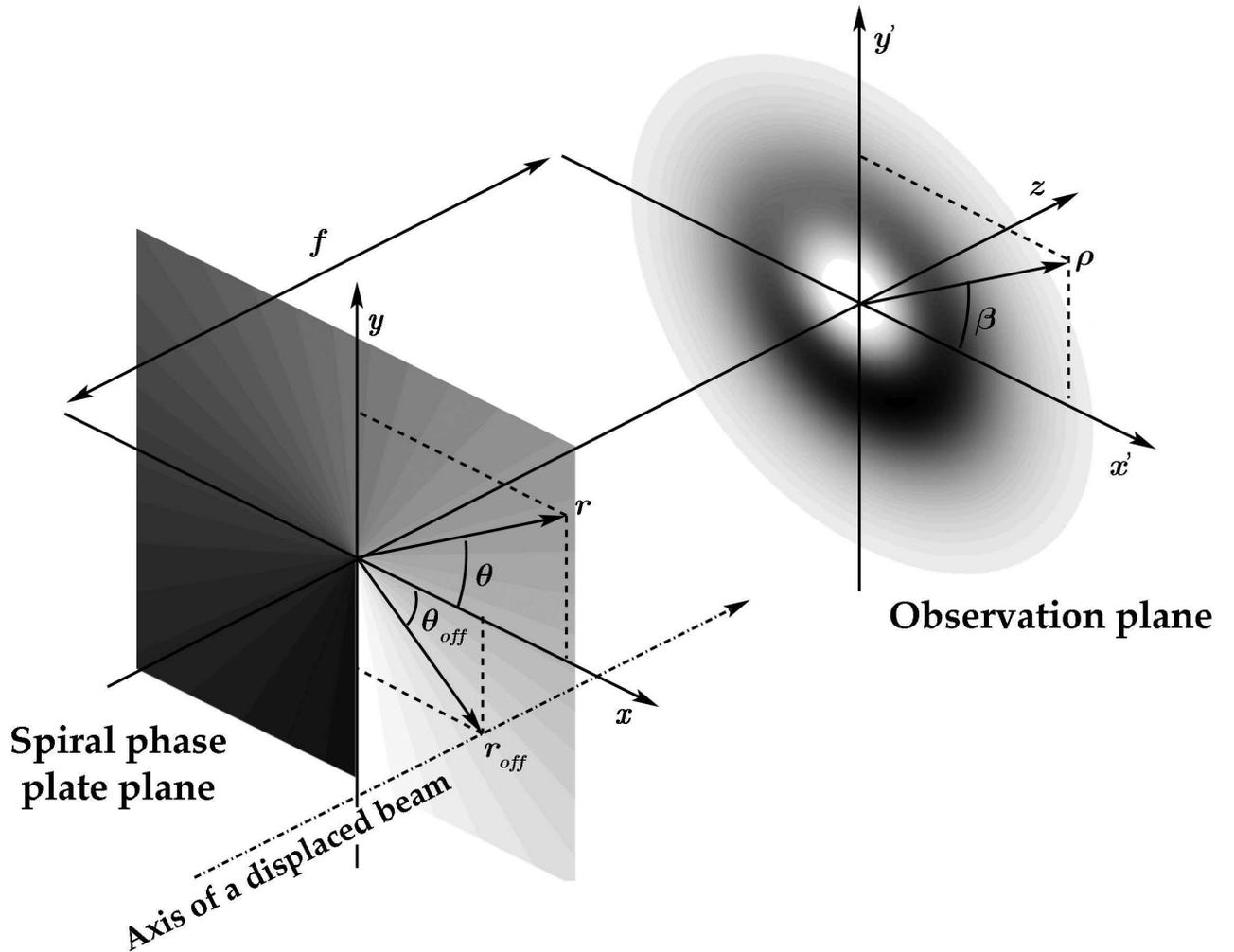}
\end{center}
\caption{The geometrical configuration adopted to study the Fraunhofer diffraction of
a Gaussian beam beyond a spiral phase plate (see text for details).}
\label{fig:geometry}
\end{figure}

In this Section we revisit the problem of the Fraunhofer diffraction of a monochromatic
Gaussian beam intersecting an SPP exactly on-axis \footnote{Similar results can be
found also for a FH, provided that the angle subtended by the first diffraction order
is small \cite{bek08a}.}. The geometrical configuration adopted here is sketched in
Fig. \ref{fig:geometry}. The SPP is placed in the $(x, y)$ plane and the central
optical singularity coincides with the origin of the $(x, y)$ coordinate system. To
take advantage of the circular symmetry of the geometry, in the following we will use
the circular coordinates $\left(r, \theta\right)$ defined by $x = r \cos \theta$ and
$y = r \sin \theta$. Thus, the transmission function of the SPP is a complex function
of the azimuthal angle $\theta$:
\begin{equation}
T_{\ell} (\theta) = \ee{\ii \, \ell \, \theta}~,
\end{equation}
where $\ell$ represents the strength of the optical singularity. We consider a Gaussian
beam propagating along the $z$ axis (that is also its symmetry axis) perpendicular to
the SPP plane. This choice is important for practical applications, i.e. laser beams or
starlight beams affected by atmospheric turbulence. We then assume that the field
amplitude distribution in the SPP plane is
\begin{equation}
A_G (r) = c \, \ee{- r^2 / w^2}~,
\end{equation}
where $c$ is a complex factor, eventually dependent on the $z$ coordinate, and $w$ is
related to the beam width. The observation plane $(x',y')$ is located at a distance
$f \rightarrow \infty$ beyond the SPP or, better, at the focal plane of a lens placed
just behind the SPP (in this case, $f$ would coincide with the focal length of the
lens). The scalar field of the beam in the observation plane can be obtained from the
Fourier transform of the product between the amplitude of the input beam and the SPP
transmission function:
\begin{equation}
\label{eqn:fraun} 
u_{\ell}(\rho, \beta) = \text{const} \iint A_G(r) \, T_{\ell}(\theta) \,
\ee{-\ii k r \rho \cos(\theta - \beta) / f} \, r \dd r \dd \theta~,
\end{equation}
where $k = 2 \pi / \lambda$ is the wave vector, $\lambda$ is the wavelength and
$(\rho, \beta)$ are the circular coordinates in the observation plane. To simplify
the calculations, the SPP is considered as infinitely extended in the $(x, y)$
plane. In addition, we use the scaled radial coordinate $r' = k r / f$ and
introduce the quantities $w_0 = f / (k w)$ and $c_0$, the latter containing all the
multiplicative constants. In this way, Eq. \ref{eqn:fraun} becomes
\begin{equation}
u_{\ell}(\rho, \beta) = c_0 \int_0^{2 \pi} \int_0^\infty \ee{-w_0^2 r'^2} \,
\ee{\ii \, \ell \, \theta} \, \ee{-\ii r' \rho \cos(\theta - \beta)}
r' \dd r' \dd \theta~.
\end{equation}
The integral involving the angular coordinate $\theta$ can be evaluated by using the
definition of the Bessel function of the first kind $\mathrm{J}_n (z)$. Thus, the
integral involving the spatial coordinate becomes a particular case of the Weber-Sonine
formula \cite{abr64}. The final result of the integration can be expressed in
terms of the confluent hypergeometric function of the first kind. However, a more
useful expression of the amplitude distribution of the output beam is obtained
by using the modified Bessel function of the first kind $\mathrm{I}_\nu (z)$.
By introducing the quantity $\eta = \rho / (2 w_0)$, the final result is \cite{sac98}
\begin{equation}
\label{eqn:kummerov1}
u_{\ell}(\rho, \beta) = c_0 \, \ii^{-\ell} \, \frac{\pi^{3/2}}{2 w_0^2}
\, \ee{\ii \, \ell \, \beta} \ee{-\eta^2 / 2} \eta
\left[\mathrm{I}_{\frac{\ell - 1}{2}}\left(\frac{\eta^2}{2}\right) -
\mathrm{I}_{\frac{\ell + 1}{2}}\left(\frac{\eta^2}{2}\right)\right]~.
\end{equation}
The presence of the phase factor $\ee{\ii \, \ell \, \beta}$ implies that the output
beam has an $\ell$-charged OV nested inside.

Beams of this kind, also known as `Kummer beams' \cite{bek08a}, are different
from the commonly used Laguerre-Gaussian (L-G) beams \cite{tur96,arl98}. If
$\ell = 0$, the Bessel functions of half-integer index in Eq. \ref{eqn:kummerov1}
can be expressed in terms of the hyperbolic functions and combined together to
give an exponential. In this case, the amplitude distribution of the output beam
is still Gaussian, i.e. $u_0 \sim \ee{-\eta^2}$. If $\ell \neq 0$, we can derive an
useful approximation for $\eta \rightarrow 0$ by using the series expansion of
$\mathrm{I}_\nu (z)$ \cite{abr64}:
\begin{equation}
\mathrm{I}_\nu (z) = {\left(\frac{z}{2}\right)}^\nu \sum_{m = 0}^\infty
\frac{{(z / 2)}^{2 m}}{m! \, \Gamma(\nu + m + 1)}~.
\end{equation}
We can recognize that, near the $z$ axis, the amplitude of an on-axis Kummer beam
carrying an OV with topological charge $\ell$ could be represented by a superposition
of amplitudes of L-G modes with $p = 0$:
\begin{equation}
u_{\ell} \sim \ee{-\eta^2 / 2} \eta^\ell
\left[\frac{1}{\Gamma\left(\frac{\ell + 1}{2}\right)} -
\frac{\eta^2}{2^2 \Gamma\left(\frac{\ell + 3}{2}\right)} +
\frac{\eta^4}{2^4 \Gamma\left(\frac{\ell + 3}{2}\right)} -
\frac{\eta^6}{2^6 \Gamma\left(\frac{\ell + 5}{2}\right)} + \mathcal{O}(\eta^8)
\right]~.
\end{equation}
The dominant term is represented by an L-G mode with index $\ell$, while higher order
terms are L-G modes with indices $\ell + 2 m$ ($m = 1, 2, \ldots$).

\subsection{Properties of the intensity distribution}
\label{sec:intensity}

\begin{figure}
\begin{center}
\includegraphics[width=\columnwidth]{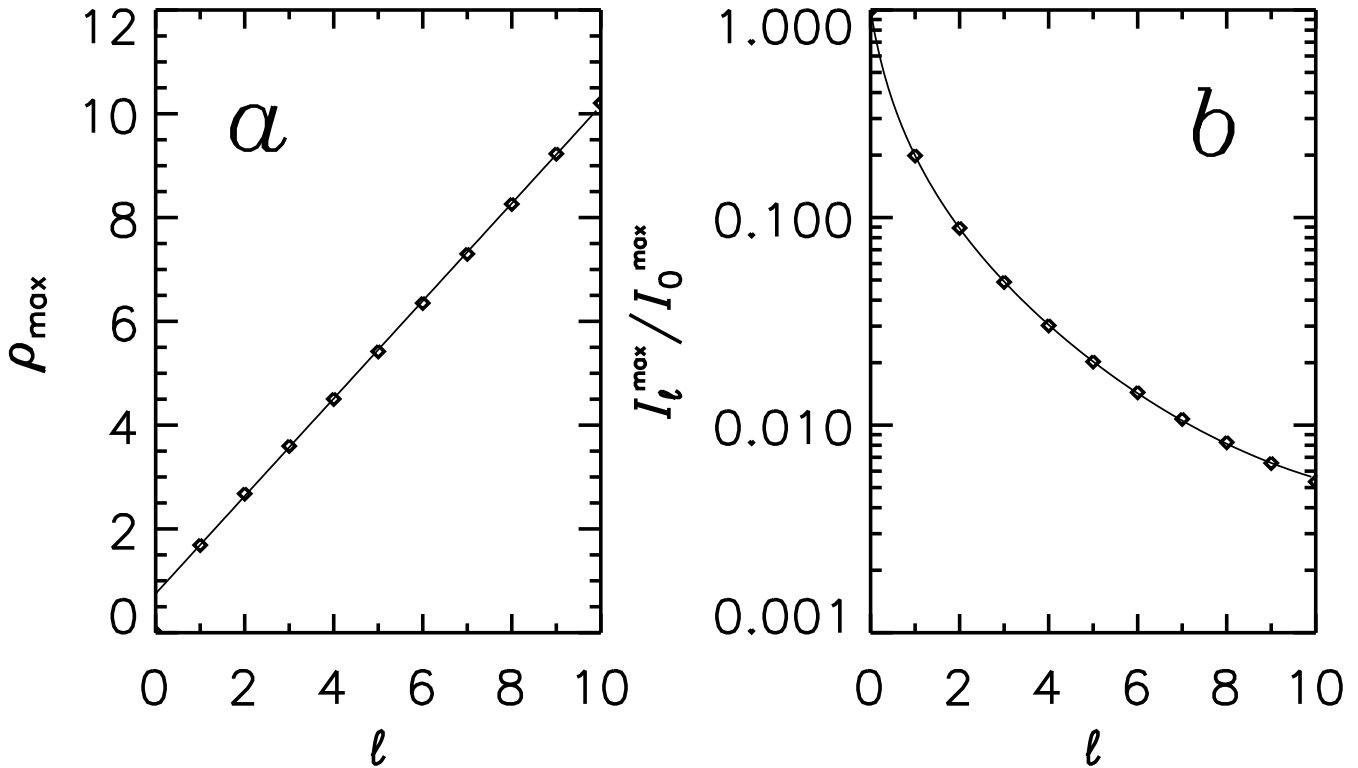}
\end{center}
\caption{Properties of the intensity distribution of on-axis Kummer beams having
$\ell = 0, 1, \ldots, 10$. (\textit{a}) Plot of the radius of maximum intensity
$\rho_{\max}$ (in units of $w_0$) vs. $\ell$. (\textit{b}) Plot of the intensity
calculated at $\rho_{\max}$ for different values of $\ell$. The maximum intensity
for $\ell = 0$ has been set to unity.}
\label{fig:kov_int}
\end{figure}

The intensity distribution of an on-axis Kummer beam is axially symmetric around
the $z$ axis and is described by:
\begin{equation}
\label{eqn:kummerov2}
I_\ell (\rho, \beta) \approx {\left|u_{\ell}(\rho, \beta)\right|}^2
= c_0^2 \, \frac{\pi^3}{4 w_0^4} \ee{-\eta^2} \eta^2
{\left[\mathrm{I}_{\frac{\ell - 1}{2}}\left(\frac{\eta^2}{2}\right) -
\mathrm{I}_{\frac{\ell + 1}{2}}\left(\frac{\eta^2}{2}\right)\right]}^2~.
\end{equation}
As for L-G modes, the intensity pattern of a Kummer beam has an annular shape,
with a central dark region where the intensity is zero. However, there
are some fundamental differences between the two analytical descriptions.
For a Kummer beam the behavior of the intensity at large distances from the
$z$ axis is $\sim \eta^{-4}$, while for an L-G mode it decreases exponentially.
Moreover, the radius of maximum intensity of an L-G mode is $\rho_{\max} \sim
\sqrt{\ell / 2}$, where the intensity attains the value $I(\rho_{\max}) \sim
\ell^\ell \, \ee{-\ell} / \ell!$, while for a Kummer beam $\rho_{\max}$ is found by
numerically solving the transcendental equation
\begin{equation}
\left(\ell + 2 \eta^2\right)
\mathrm{I}_{\frac{\ell + 1}{2}}\left(\frac{\eta^2}{2}\right) + \left(\ell -
2 \eta^2\right) \mathrm{I}_{\frac{\ell - 1}{2}}\left(\frac{\eta^2}{2}\right) = 0~.
\end{equation}
The calculation of the radii of maximum intensity obtained for a set of values
of the topological charge $\ell = 0, 1, \ldots, 10$ (see Fig. \ref{fig:kov_int}a)
suggests that $\rho_{\max}$ is linearly dependent on $\ell$:
\begin{equation}
\frac{\rho_{\max}}{2 w_0} = (0.37 \pm 0.01) + (0.470 \pm 0.002) \ell~.
\end{equation}
A similar result was found also for OVs produced by a plane wave intersecting
a finite circular phase mask \cite{cur03}. Fig. \ref{fig:kov_int}b shows the
intensity calculated at $\rho_{\max}$ for the same set of topological charges.

\subsection{The effects of off-axis displacements}
\label{sec:offaxis}

\begin{figure}
\begin{center}
\includegraphics[width=\columnwidth]{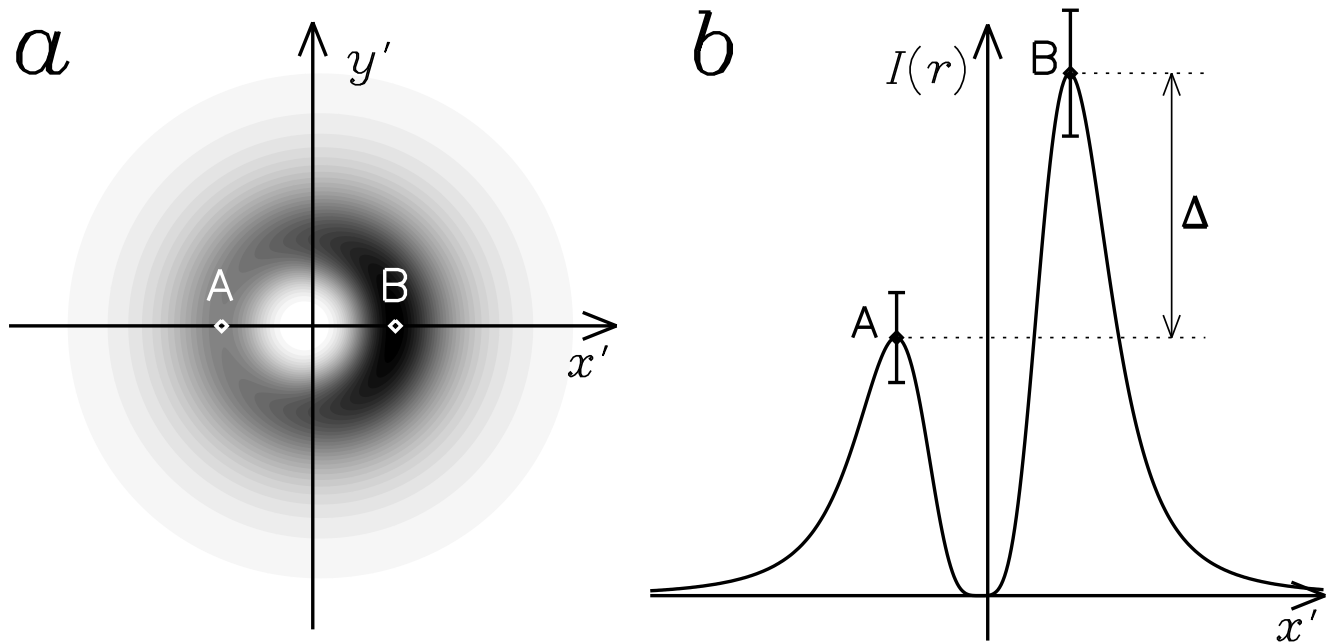}
\end{center}
\caption{Example of the far-field intensity pattern of a numerically-simulated
off-axis vortex beam produced by an $\ell = 2$ SPP. The two diamonds indicate the
positions of the two different intensity peaks $A$ and $B$. (\textit{a}) Contour plot
of the intensity distribution obtained in the observation plane. (\textit{b})
Intensity profile of the off-axis beam extracted along the $x'$ direction. $\Delta$
is the difference between the intensities calculated in $B$ and $A$. Examples of error
bars are reported for both the intensity peaks (see text).}
\label{fig:ovex}
\end{figure}

When the input Gaussian beam is displaced off-axis, so that its symmetry axis does
not coincide with the $z$ axis of Fig. \ref{fig:geometry}, the intensity pattern
produced in the observation plane is modified. The misalignment of the input beam can
be decomposed into a translation in the $(x, y)$ plane and an inclination angle
$\omega$ with respect to the $z$ axis. However, if $\omega$ is small, the modifications
induced in the intensity pattern of the output beam are negligible \cite{bek08b}. For
this reason, in our calculations we will consider only lateral displacements of the
incident beam. Let us then assume that the Gaussian beam intersects the SPP in the
position $\left(r_\text{off}, \theta_\text{off}\right)$, as shown in Fig.
\ref{fig:geometry}. The field of the output beam resembles that of Eq.
\ref{eqn:kummerov1} obtained under on-axis conditions:
\begin{equation}
\label{eqn:a_off1} 
u_{\ell}(\rho, \beta) = c_0 \, \ii^{-\ell} \frac{\pi^{3/2}}{2 w_0^2}
\ee{\ii \, \ell \, \psi} \ee{-w_0^2 {r'_\text{off}}^2} \ee{-\gamma^2 / 8 w_0^2}
\frac{\gamma}{2 w_0}  \left[\mathrm{I}_{\frac{\ell - 1}{2}}
\left(\frac{\gamma^2}{8 w_0^2}\right) - \mathrm{I}_{\frac{\ell + 1}{2}}
\left(\frac{\gamma^2}{8 w_0^2}\right)\right]~.
\end{equation}
Here, $r'_\text{off}$ is the scaled radial coordinate obtained from $r_\text{off}$,
while the quantities $\gamma$ and $\psi$ are defined as \cite{bek08b}
\begin{equation}
\label{eqn:a_off2}
\left\{
\begin{array}{lll}
\gamma^2 & = & \rho^2 + 4 \ii w_0^2 r'_\text{off} \rho \cos(\beta -
\theta_\text{off}) - 4 w_0^4 {r'_\text{off}}^2 \\
\tan \psi & = & \dfrac{\rho \sin \beta + 2 \ii w_0^2 r'_\text{off} \sin
\theta_\text{off}}{\rho \cos \beta + 2 \ii w_0^2 r'_\text{off} \cos
\theta_\text{off}}
\end{array}
\right.~.
\end{equation}
In this case, the additional exponential factor and the complex value of $\gamma^2$
produces a phase singularity which is located neither on the beam axis, nor in the
origin of the $(x', y')$ plane, but shifted in a position $\left(\rho, \beta\right) =
\left(2 w_0^2 r'_\text{off}, \theta_\text{off} + \pi / 2\right)$. As a result, the
intensity distribution of the output beam becomes asymmetric \cite{bas04}, showing two
different peaks along the direction of the vortex core in the $(x',y')$ plane.
Fig. \ref{fig:ovex}a shows an example of an off-axis OV produced with an $\ell = 2$
SPP. The lower and the higher peaks are labeled with $A$ and $B$, respectively.

Now, since it is difficult to find analytical solutions of Eqs. \ref{eqn:a_off1} and
\ref{eqn:a_off2}, we decided to perform numerical simulations. The width $w$ of the
input Gaussian beam was parametrized in function of the full width at half of the
intensity maximum, $2 a$, such that $a = w \sqrt{\ln 2 / 2}$. We used values of the
topological charge induced by the SPP in the range $\ell = 0, 1, \ldots, 10$, since
with higher values we get misleading results using the two-dimensional Fast Fourier
Transform algorithm. For each $\ell$, we considered a number of off-axis displacements
$r_\text{off} / a$ of the input beam ranging from 0 to 1 and computed the intensity
patterns of the resulting beams.

We checked the consistency of our numerical simulations by comparing them to the
analytical models (Eq. \ref{eqn:a_off1}) for a number of values of $\ell$ and off-axis
positions. To this aim, we previously normalized the intensities of both the simulated
and the analytical patterns to the corresponding maximum values. Therefore, the $B$
peak always has a normalized intensity equal to one (obviously, both the peaks $A$ and
$B$ will have the same unity intensity if $r_\text{off} / a = 0$). The residuals of the
subtraction of the theoretical intensity patterns from the simulated ones are typically
within $10^{-4}$ for positions close to peaks $A$ and $B$. We will then assume this
quantity as the intrinsic error of our numerical simulations.

\begin{figure}
\begin{center}
\includegraphics[width=\columnwidth]{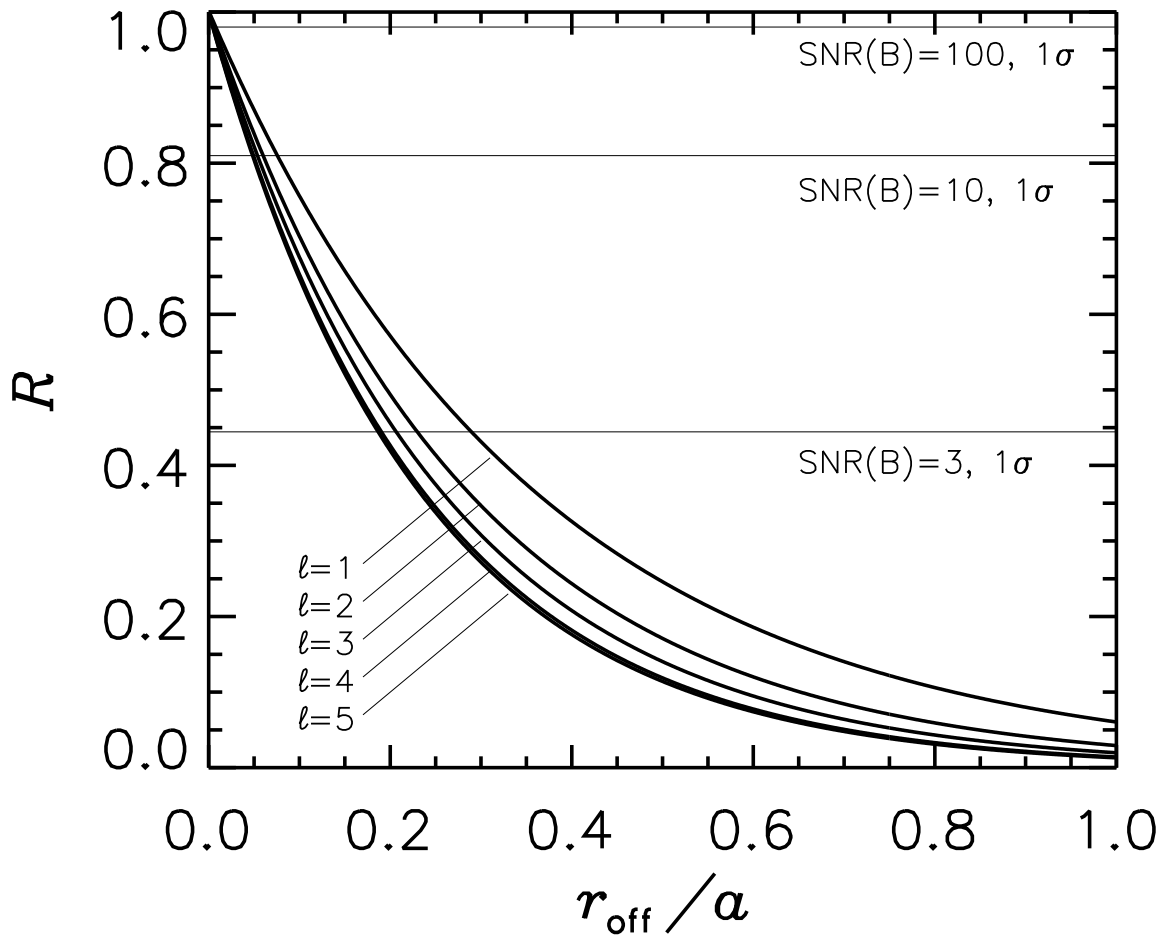}
\end{center}
\caption{Plot of the peaks intensity ratio $R$ vs. the off-axis displacement of the
input Gaussian beam obtained for different values of the topological charge induced by
the SPP. Horizontal lines are drawn at the maximum values of $R$ detectable at the
$1 \sigma$ confidence level for 3 values of the SNR ratio of the $B$ peak.}
\label{fig:ratios}
\end{figure}

For all the numerically simulated OVs we obtained the intensity values at the two peaks
and calculated the quantity $R$ defined as the ratio between the intensity $I_A$ of the
lower peak and the intensity $I_B$ of the higher peak. We find that $R$ rapidly
decreases as the off-axis displacement increases for all the topological charges
considered. The graphs showing the dependence of $R$ on $r_\text{off} / a$ for $\ell =
1, 2, 3, 4, 5$ are plotted in Fig. \ref{fig:ratios}. All the curves are well
represented by a simple exponential function:
\begin{equation}
\label{eqn:rroff}
R = k_1 \ee{-k_2 \, r_\text{off} / a}~,
\end{equation}
where parameters $k_1$ and $k_2$, obtained by best fitting the simulated curves, are
listed in Table \ref{tab:bestfit}. From these results, it appears that $k_2$ depends on
the topological charge $\ell$ as
\begin{equation}
k_2 = (4.64 \pm 0.05) - (2.9 \pm 0.2) \ee{-(0.47 \pm 0.05) \ell}~,
\end{equation}
while $k_1$ seems to remain equal to unity.

\begin{table}
\caption{Best-fit values of parameters $k_1$ and $k_2$ in Eq. \ref{eqn:rroff} obtained
from least-square fits of the simulated curves of Fig. \ref{fig:ratios}. The associated
errors are given at the $1 \sigma$ confidence level.}
\label{tab:bestfit}
\begin{ruledtabular}
\begin{tabular}{ccc}
$\ell$	& $k_1$					& $k_2$	\\
\hline
1		& $1.002 \pm 0.001$		& $2.808 \pm 0.006$	\\
2		& $1.0012 \pm 0.0007$	& $3.534 \pm 0.004$	\\
3		& $1.0004 \pm 0.0005$	& $3.928 \pm 0.003$	\\
4		& $0.9995 \pm 0.0008$	& $4.259 \pm 0.005$	\\
5		& $0.9990 \pm 0.0009$	& $4.338 \pm 0.006$	\\
6		& $0.999 \pm 0.001$		& $4.432 \pm 0.008$	\\
7		& $0.998 \pm 0.001$		& $4.456 \pm 0.008$	\\
8		& $0.998 \pm 0.001$		& $4.509 \pm 0.009$	\\
9		& $0.998 \pm 0.001$		& $4.573 \pm 0.01$	\\
10		& $0.998 \pm 0.002$		& $4.756 \pm 0.02$	\\
\end{tabular}
\end{ruledtabular}
\end{table}

\section{The sensitivity of the method to reveal off-axis displacements}
\label{sec:ratio}

The off-axis displacement of the input Gaussian beam with respect to the central
singularity of an SPP results in an asymmetry of the far-field intensity pattern.
Eq. \ref{eqn:rroff} reveals that the parameter $R$ represents an extremely sensitive
tool to detect such very small displacements. 

Let us suppose to have an input Gaussian beam which symmetry axis is perpendicular
to the surface of an SPP and observe the correspondent far-field image with a
photoelectric detector like a CCD. In this way, if the beam is displaced off-axis, we
can measure the intensity ratio $R$ of the two different peaks, as defined in Sect.
\ref{sec:offaxis}. The precision of this measurement is mainly limited by the
signal-to-noise ratio (SNR) achieved in the observation, while additional errors might
be introduced by construction imperfections of the SPP. The latter issue results in
inhomogeneities of the observed intensity distribution. For this reason, efforts are
currently made to improve the production quality of SPPs \cite{oem04b,wat04,sue04}.
However, here we assume to use an ideal SPP so that the only limitations are due to
the SNR. The noise of a CCD detector is mainly represented by the photon shot noise
\cite{how06}. Assuming a pure Poissonian distribution of the collected photons, the
uncertainties associated to the intensities $I_A$ and $I_B$ of the two peaks can be
approximated by the square root of the signals, i.e. $\sigma_A = \sqrt{I_A}$ and
$\sigma_B = \sqrt{I_B}$. We may then recognize $I_A \neq I_B$ at the $n \sigma$
confidence level when $\Delta = I_B - I_A \geq n \sigma_A + n \sigma_B$ (see Fig.
\ref{fig:ovex}b), that means
\begin{equation}
\Delta \geq n \left(\sqrt{I_A} + \sqrt{I_B}\right)~.
\end{equation}
If we introduce parameter $R$, this equation can be rewritten as a function of the only
SNR associated to the intensity of the highest peak ($\text{SNR}(B)$). We find that
the maximum peaks intensity ratio measurable at the $n \sigma$ confidence level is
\begin{equation}
\label{eqn:snr}
R \leq {\left(1 - \frac{n}{\text{SNR}(B)}\right)}^2~.
\end{equation}
As useful examples, in Fig. \ref{fig:ratios} we draw three horizontal lines
corresponding
to the maximum $R$ values 0.44, 0.81 and 0.98 detectable at the $1 \sigma$ level for
$\text{SNR}(B) = 3$, 10 and 100, respectively. By combining Eq. \ref{eqn:rroff} with
Eq. \ref{eqn:snr}, we finally obtain the expression for the minimum off-axis
displacement detectable at the $n \sigma$ confidence level:
\begin{equation}
\frac{r_\text{off}}{a} \geq -\frac{1}{k_2} \ln \left[\frac{1}{k_1}
{\left(1 - \frac{n}{\text{SNR}(B)}\right)}^2\right]~.
\end{equation}

One general outcome is that, for a fixed $\text{SNR}(B)$, OVs with higher $\ell$ values
allow the detection of smaller off-axis displacements. This effect is more significant
at low SNR regimes, when the maximum measurable $R$ is small and the curves in Fig.
\ref{fig:ratios} are more spatially separated. As $\text{SNR}(B)$ increases, the
advantage obtained by using high values of the topological charge becomes negligible.
In fact, if we assume $\text{SNR}(B)$ above 10, we might reveal off-axis displacements
$< 0.1 a$ for all $\ell$ values. Instead, considering the lowest acceptable value
$\text{SNR}(B) = 3$ for signal detection, we can detect off-axis displacements of
$\sim 0.3 a$ for $\ell = 1$ at the $1 \sigma$ confidence level.

\section{Conclusions}
\label{sec:conclusions}

In this Paper we have analyzed the properties of the Fraunhofer diffraction pattern
produced by a Gaussian light beam crossing an SPP. When the input beam is perfectly
aligned with the central singularity of the SPP, the resulting beam is a Kummer beam
with a symmetric annular intensity distribution. Instead, an off-axis displacement of
the input beam produces an asymmetry in the far-field intensity pattern. In particular,
the intensity profile along the direction of maximum asymmetry shows two different
peaks. We have found that, for all the values of the topological charge considered,
the ratio $R$ of their intensities changes exponentially with the off-axis displacement
of the input beam. We have quantitatively analyzed how the SNR associated to the
highest peak affects the sensitivity of the ratio $R$ in revealing very small
misalignments of the input beam. In particular, we have found that higher values of
the topological charge $\ell$ generally provide better resolutions, especially for low
SNR regimes. We suggest that this method could find interesting applications in
high-precision positioning systems. Note that similar results can be obtained also
by using other spatial properties of transverse laser modes \cite{tre03}. Finally,
the sensitivity of OVs could be used in astrometry, by placing an SPP at the focal plane
of a telescope.

\begin{acknowledgments}
We gratefully acknowledge financial support from the CARIPARO foundation.
\end{acknowledgments}

\bibliographystyle{apsrev}
\bibliography{biblio}

\end{document}